\documentclass[journal,onecolumn,web]{article}
\usepackage{mathtools}
\usepackage{arxiv}
\usepackage{cite}
\usepackage{amsmath,amssymb,amsfonts}
\usepackage{amsthm}
\usepackage{algorithmic}
\usepackage{graphicx}
\newtheorem{theorem}{Theorem}[section]

\newtheorem{assumption}[theorem]{Assumption}
\newtheorem{corollary}[theorem]{Corollary}

\newtheorem{definition}[theorem]{Definition}
\newtheorem{lemma}[theorem]{Lemma}
\newtheorem{remark}[theorem]{Remark}

\newcommand{\R}{{\mathbb{R}}}
\newcommand{\N}{{\mathbb{N}}}

\usepackage{hyperref}
\hypersetup{hidelinks=true}
\usepackage{textcomp}
\def\BibTeX{{\rm B\kern-.05em{\sc i\kern-.025em b}\kern-.08em
    T\kern-.1667em\lower.7ex\hbox{E}\kern-.125emX}}
\begin{document}
\title{Backstepping Design for Incremental Input-to-State Stabilization of Unknown Systems}
\author{David Smith Sundarsingh, Bhabani Shankar Dey, and Pushpak Jagtap
\thanks{This work was supported in part by the SERB Start-up Research Grant and the ARTPARK.}
\thanks{D. S. Sundarsingh, B. S. Dey, and P. Jagtap are with Robert Bosch Centre for Cyber-Physical Systems, IISc, Bangalore, India {\tt\small david\_smith\_16@outlook.com, bhabanishankar440@gmail.com, pushpak@iisc.ac.in}}%
}
\maketitle
\begin{abstract}
Incremental stability of dynamical systems ensures the convergence of trajectories from different initial conditions towards each other rather than a fixed trajectory or equilibrium point. Here, we introduce and characterize a novel class of incremental Lyapunov functions, an incremental stability notion known as Incremental Input-to-State practical Stability ($\delta$-ISpS). Using Gaussian Process, we learn the unknown dynamics of a class of control systems. We then present a backstepping control design scheme that provides state-feedback controllers that render the partially unknown control system $\delta$-ISpS. To show the effectiveness of the proposed controller, we implement it in two case studies.
\end{abstract}
\section{Introduction}
A stronger stability property of non-linear systems called Incremental Stability ensures the convergence of trajectories towards each other rather than to a specific trajectory or equilibrium point. This notion has recently been extensively studied due to its applicability in the synchronization of cyclic feedback systems \cite{synch}, complex networks \cite{synchComplex} and interconnected oscillators \cite{synchOsci}, modelling of nonlinear analog circuits \cite{modelCirc}, and symbolic model construction for nonlinear control systems \cite{bisim1,bisim2,zamani2017towards,jagtap2020symbolic,jagtap2017quest}.

Incremental Input-to-State Stability ($\delta$-ISS), a particular class of incremental stability, has been extensively studied and characterized by Lyapunov functions \cite{characterize1,angeli,zamanicharacterize}. In addition, state feedback controllers for rendering a class of control systems $\delta$-ISS have been designed. Examples include works on unstable non-smooth control systems \cite{zamaninonsmooth}, stochastic systems \cite{pushpakHamilton} and backstepping approach \cite{deltaISS,backsteppingzamani}. While controller synthesis for $\delta$-ISS stabilization has been studied extensively, to the best of the authors' knowledge, there is no work on controller synthesis for an unknown system. We aim to address this problem by learning the unknown system dynamics by using the Gaussian Process and developing a backstepping control design scheme based on the learned system model. We consider a class of partially unknown control systems represented in the strict feedback form.

Gaussian process (GP) has been used for system identification in various works in the literature due to its ability to approximate unknown nonlinear dynamics while providing a measure of the model fidelity \cite{GPBook}. It has been used in works on tracking control \cite{tracking}, feedback linearization \cite{feedbackLinearization}, control Lyapunov approach \cite{hircheControl}, and control barrier functions \cite{JagtapGP}. Since GP-based system models are just approximations, it is not possible to ensure strict $\delta$-ISS by using the learned models. But, it is possible to ensure relaxation of the property called incremental Input-to-State practical Stability ($\delta$-ISpS). We define and characterize this notion {for the first time} in this paper based on the notion of Input-to-State practical Stability introduced in \cite{ISpS}. We then use a backstepping control design scheme to synthesize controllers that ensure $\delta$-ISpS property. 

In this paper, we present the definition and characterization of Incremental Input-to-Space practical Stability. We use Gaussian Process to learn the unknown dynamics of the partially unknown system, given in strict-feedback form, using the methodology introduced in \cite{GPBackStepping}. We then provide a backstepping control design scheme along with the corresponding $\delta$-ISpS Lyapunov functions based on a filtered command backstepping approach that synthesizes controllers for rendering the system $\delta$-ISpS. To the best of the authors' knowledge, this is the first work that synthesizes a controller for guaranteed (probabilistic) incremental stabilization of a class of partially unknown systems given in a strict feedback form. To show the practicality of the approach, we implement the controllers synthesized based on the proposed design scheme in two case studies to show that the trajectories of the system do indeed converge.
\section{Incremental Input-to-State practical Stability}
\subsection{Notations}
The set of real, positive real, non-negative real and positive integers are given by $\R$, $\R^+$, $\R^+_0$ and $\N$, respectively. $\R^n$ denotes an $n$-dimensional Euclidean space and $\R^{n\times m}$ represents the space of real valued matrices with $n$ rows and $m$ columns. Given a vector $x\in\R^n$, $x_i$ denotes its $i^{th}$ element, $\lVert x\rVert=\max\left\{|x_1|,\ldots,|x_n|\right\}$, its infinity norm and $|x_i|$ is the absolute value of $x_i$. Given a measurable function $\upsilon:\R^+_0\rightarrow\R^n$, $\lVert\upsilon\rVert_{\infty}:=(ess)sup\left\{\lVert\upsilon(t)\rVert,t\geq 0\right\}$ is its (essential) supremum. $\mathbf{I}_n\in\R^{n\times n}$ and $0_n\in\R^n$ represents identity matrix and zero vector respectively. $\mathcal{G}(\mu,C)$ denotes the multivariate gaussian distribution, where $\mu\in\R^n$ is the mean vector and $C\in\R^{n\times n}$ is the covariance matrix. The reproducing kernel Hilbert space (RKHS) is a Hilbert space of square-integrable functions equipped with an RKHS norm denoted by $\lVert f\rVert_{k}$, where $f$ is a function, $k:\mathcal{X}\times \mathcal{X}\rightarrow\R^+_0$ is a symmetric positive definite function referred to as a kernel and $\mathcal{X}\subset\R^n$. Note that RKHS includes functions of the form $f(x)=\Sigma_ia_ik(x,x_i)$, where $a_i\in\R$, $x,x_i\in \mathcal{X}$ and $k$ is a kernel. A detailed discussion on RKHS and RKHS norms can be found in \cite{RKHS}. A continuous function $\alpha:\R^+_0\rightarrow\R^+_0$ is class-$\mathcal{K}$ if $\alpha(0)=0$ and if it is strictly increasing. If $\alpha\in\mathcal{K}$ is unbounded, i.e $\alpha(r)\rightarrow\infty$ as $r\rightarrow\infty$, $\alpha\in\mathcal{K}_{\infty}$. A continuous function $\beta:\R^+_0\times\R^+_0\rightarrow\R^+_0$ belongs to class-$\mathcal{KL}$ if for a fixed $s$, $\beta(r,s)\in\mathcal{K}_{\infty}$ with respect to $r$ and for fixed $r$, $\beta(r,s)$ is decreasing with increase in $s$ and $\beta(r,s)\rightarrow 0$ as $s\rightarrow\infty$. In this paper, we consider a class of non-linear systems with $h\in\N$ subsystems expressed in strict-feedback form \cite{strictFeedback}. For any $x,y,z\in\R^{d}$, $\mathbf{d}:\R^{d}\times\R^{d}\rightarrow\R_0^+$ is a metric on $\R^{d}$ if: $(i)$ $\mathbf{d}(x,y)=0$, iff $x=y$; $(ii)$ $\mathbf{d}(x,y)=\mathbf{d}(y,x)$; $(iii)$ $\mathbf{d}(x,z)\leq\mathbf{d}(x,y)+\mathbf{d}(y,z)$.
\subsection{Control System}
Consider a class of control systems defined as follows:
\begin{definition}\label{def:sys1}
    A control system is a quadruple $\Sigma = (\mathcal{X},U,\mathcal{U},f)$, where
    \begin{itemize}
        \item $\mathcal{X}\subseteq\R^d$ is the state space;
        \item $U\subseteq\R^m$ is the input space; 
        \item $\mathcal{U}$ is the subset of all measurable functions of time with values in $U$;
        \item $f:\mathcal{X}\times U\rightarrow\R^d$ is a map satisfying the local Lipschitz continuity assumption. This assumption ensures the existence and uniqueness of trajectories \cite{existence}.
    \end{itemize}
    A curve $\xi:\R^+_0\rightarrow\R^{d}$ is said to be the trajectory of $\Sigma$ if there exists $\upsilon\in\mathcal{U}$ such that:\begin{gather}\label{equ:sys1}
        \dot\xi=f(\xi,\upsilon).
    \end{gather} 
\end{definition}
 We use the notation $\xi_{x\upsilon}(t)$ to denote the value along the trajectory that is reached at time $t$ under the input signal $\upsilon\in\mathcal{U}$ from the initial state $x=\xi_{x\upsilon}(0)$. It is assumed that state space $\mathcal{X}$ is forward invariant under \eqref{equ:sys1}, i.e., $x\in\mathcal{X}\implies \xi_{xv}(t)\in \mathcal{X}$ for all $t \geq 0$.
\subsection{Incremental Input-to-State Practical Stability}
In this subsection, we introduce the notion of Incremental Input-to-State practical Stability ($\delta$-ISpS) and its characterization using the $\delta$-ISpS Lyapunov Function. The stability notion presented is inspired by the notion of Input-to-State practical Stability presented in \cite{ISpS}.
\begin{definition}\label{def:deltaISpS}
    A control system $\Sigma$ is called \textit{incrementally input-to-state practically stable} ($\delta$-ISpS), if there exists a metric $\mathbf{d}$, functions $\beta\in\mathcal{KL}$, $\gamma\in\mathcal{K}_{\infty}$ and a constant $c>0$ such that for any $t\in\R^+_0$, any $x,x'\in \mathcal{X}$, and any $\upsilon,\upsilon'\in\mathcal{U}$, the following holds,
    \begin{align}\label{equ:deltaISpS}
        \mathbf{d}(\xi_{x\upsilon}(t),\xi_{x'\upsilon'}(t))\leq\beta(\mathbf{d}(x,x'),t)+\gamma(\lVert \upsilon-\upsilon'\rVert_{\infty})+c.
    \end{align}
\end{definition}It is obvious from (\ref{equ:deltaISpS}) that if $c=0$, the system is \textit{incrementally input-to-state stable} as defined in \cite{angeli}.

In order to characterize the $\delta$-ISpS property of the system, we introduce the notion of $\delta$-ISpS Lyapunov functions in the following definition.
\begin{definition}\label{def:deltaISpSLyapunov}
    Consider a control system $\Sigma$ as defined in Definition \ref{def:sys1} and a differentiable function $V:\R^{d}\times\R^{d}\rightarrow\R^+_0$. Function $V$ is called a $\delta$-ISpS Lyapunov function if there exist functions $\overline{\alpha},\underline{\alpha},\sigma\in\mathcal{K}_{\infty}$ and constants $\Tilde{c},k\in\R^+$, such that
    \begin{enumerate}\label{item:lyapunov1}
        \item $\forall x,x'\in \mathcal{X}$, $\underline{\alpha}(\mathbf{d}(x,x'))\leq V(x,x')\leq\overline{\alpha}(\mathbf{d}(x,x'))$;
        \item $\forall x,x'\in \mathcal{X}$ and $\forall u,u'\in {U}$, the following holds: $\dot V(x,x')\leq-kV(x,x')+\sigma(\lVert u-u'\rVert)+\Tilde{c}$.
    \end{enumerate}
\end{definition}
The following theorem describes $\delta$-ISpS in terms of the existence of a $\delta$-ISpS Lyapunov function. 
\begin{theorem}\label{theorem:ISpSLyapunov}
    A control system $\Sigma$ is $\delta$-ISpS if it admits a $\delta$-ISpS Lyapunov function.
\end{theorem}
\begin{proof}
    This proof is inspired by that of \cite[Theorem 3.3]{deltaISSProof}. Consider a Lyapunov function satisfying the conditions $(1)$ and $(2)$ in Definition \ref{def:deltaISpSLyapunov}. For any $t\in\R^+_0$, $\upsilon,\upsilon'\in\mathcal{U}$ and $x,x'\in \mathcal{X}$,
    \begin{align}
        V(\xi_{x\upsilon}(t),\xi_{x'\upsilon'}(t))=V(x,x')+\int_0^t\dot V(x,x')ds.
    \end{align}
    Due to condition $(2)$ in Definition \ref{def:deltaISpSLyapunov},
    \begin{align*}
        V(\xi_{x\upsilon}(t),\xi_{x'\upsilon'}(t))&\leq\hspace{-0.2em} V(x,x'\hspace{-0.1em})\hspace{-0.2em} +\hspace{-0.4em} \int_0^t\hspace{-0.7em} \left(\hspace{-0.2em}-kV(\xi_{x\upsilon}(\hspace{-0.1em}s\hspace{-0.1em}),\xi_{x'\upsilon'}(\hspace{-0.1em}s\hspace{-0.1em})\hspace{-0.1em})\hspace{-0.2em} +\hspace{-0.2em} \sigma(\lVert \upsilon(s)\hspace{-0.2em} -\hspace{-0.2em} \upsilon'(s)\rVert)\hspace{-0.2em} +\hspace{-0.2em} \Tilde{c}\right)\hspace{-0.2em} ds\\
        &\leq \hspace{-0.2em}-k\int_0^t\hspace{-0.6em}V(\xi_{x\upsilon}(\hspace{-0.1em}s\hspace{-0.1em}),\xi_{x'\upsilon'}(\hspace{-0.1em}s\hspace{-0.1em}))ds\hspace{-0.2em}+\hspace{-0.2em}V(x,x')\hspace{-0.2em}+\hspace{-0.2em}t\sigma(\lVert \upsilon-\upsilon'\rVert_{\infty})\hspace{-0.2em}+\hspace{-0.2em}t\Tilde{c}.
    \end{align*}
    By applying Gronwall's inequality, one obtains
    \begin{align*}
         V(\xi_{x\upsilon}(t),\xi_{x'\upsilon'}(t))\hspace{-0.2em}&\leq \hspace{-0.2em}e^{-kt}V(x,x')\hspace{-0.2em}+\hspace{-0.2em}te^{-kt}(\sigma(\lVert \upsilon-\upsilon'\rVert_{\infty})\hspace{-0.2em}+\hspace{-0.2em}\Tilde{c}),\\
         &\leq e^{-kt}V(x,x')+\frac{1}{e k}\sigma(\lVert \upsilon-\upsilon'\rVert_{\infty})+\frac{1}{e k}\Tilde{c}.
    \end{align*}
    From the condition $(1)$ of Definition \ref{def:deltaISpSLyapunov}, we have,
    \begin{align*}
        &\notag\underline{\alpha}(\mathbf{d}(\xi_{x\upsilon}(t),\xi_{x'\upsilon'}(t)))
        \notag\leq e^{-kt}\overline{\alpha}(\mathbf{d}(x,x'))+\frac{1}{e k}\sigma(\lVert \upsilon-\upsilon'\rVert_{\infty})+\frac{1}{e k}\Tilde{c},\\
        &\mathbf{d}(\xi_{x\upsilon}(t),\xi_{x'\upsilon'}(t))\leq\underline{\alpha}^{-1}\left(e^{-kt}\overline{\alpha}(\mathbf{d}(x,x'))+\frac{1}{e k}\sigma(\lVert \upsilon-\upsilon'\rVert_{\infty})+\frac{1}{e k}\Tilde{c}\right).
    \end{align*}
    Since $\underline{\alpha}$ is a $\mathcal{K}_{\infty}$ function, $\underline{\alpha}^{-1}\in\mathcal{K}_{\infty}$ and the following holds: $\underline{\alpha}^{-1}(p+q+r)\leq\underline{\alpha}^{-1}(3\max(p,q,r)) \leq\underline{\alpha}^{-1}(3p)+\underline{\alpha}^{-1}(3q)+\underline{\alpha}^{-1}(3r)$.
    Using this inequality and substituting $\Tilde{c}=\underline{\alpha}(c)\frac{ek}{3}$, we get
    \begin{align}\label{equ:withoutSub}
        &\notag\mathbf{d}(\xi_{x\upsilon}(t),\xi_{x'\upsilon'}(t))\\ &\notag\leq\underline{\alpha}^{-1}\left(3e^{-kt}\overline{\alpha}(\mathbf{d}(x-x'))\right)+\underline{\alpha}^{-1}\left(\frac{3}{e k}\sigma(\lVert \upsilon-\upsilon'\rVert_{\infty})\right)+\underline{\alpha}^{-1}\left(\frac{3}{e k}\underline{\alpha}(c)\frac{e k}{3}\right)\notag\\
        &\leq\beta(\mathbf{d}(x,x'),t)+\gamma(\lVert \upsilon-\upsilon'\rVert_{\infty})+c
    \end{align}
    with $\beta(r,s)=\underline{\alpha}^{-1}(3e^{-ks}\overline{\alpha}(r))$ and $\gamma(r)=\underline{\alpha}^{-1}\left(\frac{3}{ek}\sigma(r)\right)$ for all $r\geq0$. This implies that the system $\Sigma$ is $\delta$-ISpS as defined in Definition \ref{def:deltaISpS}.
\end{proof}
\section{System Description}
In this paper, we consider a class of partially unknown non-linear systems with $h\in\N$ subsystems expressed in the strict-feedback form. We aim to design a backstepping control design scheme in order to enforce $\delta$-ISpS properties to the specified class of partially unknown systems, as described in this section.
\subsection{Strict Feedback-form}
We consider a class of control systems $\Sigma=(\mathcal{X},U,\mathcal{U},f)$ with $f$ expressed in strict feedback-form as formulated below:
\begin{align}\label{equ:sys}
    \notag&\dot\xi_i=f_i(\nu_i)+b_i\xi_{i+1},\forall i\in I\setminus\{h\},\\
    &\dot\xi_h=f_h(\nu_h)+b_h\upsilon,
\end{align}where $I=\{1,\ldots,h\}$, $\xi(t)=[\xi_1(t),\ldots,\xi_h(t)]^{\top}\in\mathcal{X}\subset\R^d$ is the state of the system, $\mathcal{X}=\prod_{i\in I}\mathcal{X}_i$, $\xi_i(t)\in\mathcal{X}_i\subset\R^n$, $\forall i\in I$, $\upsilon\in\mathcal{U}\subseteq\R^n$ is the input signal, $b_i\in\R$ for all $i\in I$ and $\nu_i=[\xi_1,\ldots,\xi_i]^{\top}$, $\forall i\in I$ denotes the concatenations of the states. Similarly, $\mathcal{N}_i=\prod_{j\in\{1,\ldots,i\}}\mathcal{X}_j$ is the concatenations of state sets of the state space. For brevity, whenever we use the subscript $i$, the full set $I$ is referred to unless specified otherwise. Since $f_i$ is a vector function, we use $f_{i,j}$ to represent its $j^{th}$ component, where $j\in\{1,\ldots,n\}$. The function $f_i$ is also assumed to vanish at $\xi_i(0)$, which is common for systems in strict feedback form \cite{strictFeedback}. The strict feedback form can be used to describe a wide variety of systems, and readers are directed towards \cite{strictFeedback} for a detailed discussion. We have the following assumptions about the system:
\begin{assumption}\label{assume:system}
    For system (\ref{equ:sys}), we assume that the functions $f_i$ are unknown and the constant $b_i\in\R$ is known.
\end{assumption}

In order to learn the unknown part of the dynamics ($f_i$), we will utilize the Gaussian process (GP) \cite{GPBook}. In order to use GP, we need the following assumption concerning the reproducing kernel Hilbert space (RKHS) norm $\lVert f_i\rVert_{k_i}$ with respect to a kernel $k_i:\mathcal{N}_i\times \mathcal{N}_i\rightarrow\R$.
\begin{assumption}\label{assume:RKHS}
    The function $f_i$ has a bounded reproducing kernel Hilbert space norm with respect to a known kernel $k_i$. That is $\lVert f_{i,j}\rVert_{k_i}\leq B_{f_i}<\infty$.
\end{assumption}In the space of continuous functions restricted to a compact set $\mathcal{X}_i$, the RKHS is dense for most kernels used. This allows the kernels to approximate any function in $\mathcal{X}_i$ \cite{InfoGain}.
\subsection{Gaussian Process}
Gaussian Process (GP) is a non-parametric regression tool that aims to approximate a nonlinear map $f_i:\mathcal{N}_i\rightarrow\R^n$ using potentially noisy measurements while also providing a bound on model accuracy \cite{GPBook}. In this work, we consider $\sigma$-sub-Gaussian noise as defined below:
\begin{definition}[\cite{GPBackStepping}]\label{def:noise}
    A scalar random variable $w_i$ is said to be $\sigma$-sub-Gaussian \cite{GPrandVar} if the following holds:
    \begin{align}\label{equ:noise}
        \mathbb{E}(e^{tw_i})\leq e^{\frac{\sigma^2t^2}{2}}, \forall t\in\R,
    \end{align}
    where $\sigma>0$ and $\mathbb{E}$ denotes the expected value operator.
\end{definition}
Given the definition of sub-Gaussian noise in Definition \ref{def:noise}, we have the following assumption on the availability of data for training the GP model.
\begin{assumption}\label{assume:availabledata}
    The measurements $n_i\in\mathcal{N}_i$ and $y_i=f_i(n_i)+w_i$, $i\in I$, are accessible at all times, where $w_i$ is an additive noise as shown in Definition \ref{def:noise}.
\end{assumption}

The map $f_i(n_i)$ can be practically approximated by using the state measurements obtained after running the system for a sufficiently small sampling time from various initial conditions with input signal $\upsilon\equiv0$. To accommodate the
approximation uncertainties, we use the additive noise $w_i$ \cite{JagtapGP}.

GP is denoted as $\mathcal{GP}(m,k)$ and is described by a mean function $m$ and a kernel $k$. Since $f_i$ is $n$-dimensional, each component of $f_i$ is approximated with a Gaussian process,
\begin{align}
    \hat{f}_{i,j}(n_i)\sim\mathcal{GP}(m_{i,j}(n_i),k_{i,j}(n_i,n'_i)),\forall j\in\{1,\ldots,n\},
\end{align}where $m_{i,j}:\mathcal{N}_i\rightarrow\R$ is a mean function and $k_{i,j}:\mathcal{N}_i\times\mathcal{N}_i\rightarrow\R$ is a kernel which is a measure of the similarity between any $n_i,n'_i\in \mathcal{N}_i$. Even though any real-valued function can be used for the prior mean function, it is common practice to set $m_{i,j}(n_i)=0$ for all $j\in\{1,\ldots,n\}$ and $n_i\in\mathcal{N}_i$. The kernel function, however, is problem dependent with the most commonly used kernels being linear, squared-exponential and Mat\'ern kernels \cite{GPBook}. The approximation of $f_i$ is given by $n$ independent GPs as follows,
\begin{align}
    \hat{f}_i(n_i)=\begin{cases}
        \hat{f}_{i,1}(n_i)\sim\mathcal{GP}(0,k_{i,1}(n_i,n'_i)),\\
        \hspace{1.5cm}\vdots\\
        \hat{f}_{i,n}(n_i)\sim\mathcal{GP}(0,k_{i,n}(n_i,n'_i)).
    \end{cases}
\end{align}
Given a dataset $\mathcal{D}_i=\{(n_i^{(j)},y_i^{(j)})\}_{j=1}^N$, where $y_i^{(j)}=f_i(n_i^{(j)})+w^{(j)},\forall j\in\{1,\ldots,N\}$ as defined in Assumption \ref{assume:availabledata} and an arbitrary state $n_i\in \mathcal{N}_i$, the inferred output $f_{i,j}(n_i)$ is approximated by a gaussian distribution $\mathcal{G}(\mu_{i,j}(n_i),\rho_{i,j}(n_i))$ with mean and covariance given as,
\begin{align}
    &\label{equ:mean_i}\mu_{i,j}(n_i)=\Bar{k}_{i,j}^T(K_{i,j}+\rho_{f_i}^2\mathbf{I}_N)^{-1}y_{i,j},\\
    &\label{equ:variance_i}\rho^2_{i,j}(n_i)=k_{i,j}(n_i,n_i)-\Bar{k}_{i,j}^T(K_{i,j}+\rho^2_{f_i}\mathbf{I}_N)^{-1}\Bar{k}_{i,j},
\end{align}
where $\Bar{k}_{i,j}=[k_{i,j}(n_i^{(1)},n_i),\ldots,k_{i,j}(n_i^{(N)},n_i)]^{\top}\in \R^N$, $y_{i,j}=[y_{i,j}^{(1)},\ldots,y_{i,j}^{(N)}]^{\top}\in\R^N$ and $$K_{i,j}=\begin{bmatrix}k_{i,j}(n_i^{(1)},n_i^{(1)}) & \ldots & k_{i,j}(n_i^{(1)},n_i^{(N)})\\\vdots & \ddots & \vdots\\   k_{i,j}(n_i^{(N)},n_i^{(1)}) & \ldots & k_{i,j}(n_i^{(N)},n_i^{(N)})
\end{bmatrix}\in \R^{N\times N}.$$ The dataset $\mathcal{D}_i$ can be created from the measurements of the derivatives $\dot \xi_i$, by subtracting the known quantities $b_1\xi_2,\ldots b_{h-1}\xi_h$ and $b_h\upsilon$. A bound $\Bar{\rho}^2_{i,j}=\max_{n_i\in \mathcal{N}_i}\rho^2_{i,j}(n_i)$ exists due to the continuity of the kernels. The overall function $\hat f_i(n_i)\sim\mathcal{G}(\mu_i(n_i),\rho_i(n_i))$, where
\begin{align}
    \label{equ:mean}\mu_i(n_i):=[\mu_{i,1}(n_i),\ldots,\mu_{i,n}(n_i)]^T,\\
    \label{equ:variance}\rho_i^2(n_i):=[\rho^2_{i,1}(n_i),\ldots,\rho^2_{i,n}(n_i)]^T.
\end{align}
Due to Assumption \ref{assume:RKHS}, the difference between the true $f_i(n_i)$ and the inferred mean $\mu_i(n_i)$ can be upper bounded with a high probability, as shown in the following Lemma.
\begin{lemma}\label{lemma:probBound}
    Given the system (\ref{equ:sys}) with assumptions \ref{assume:system} and \ref{assume:RKHS} and corresponding approximated GP model with mean and standard deviation given by (\ref{equ:mean}) and (\ref{equ:variance}), respectively, the following holds with probability of at least $1-\epsilon$,
    \begin{align}
        \bigcap_{i=1}^h\lVert f_i(n_i)-\mu_i(n_i)\rVert\leq\lVert\eta_i\rVert\lVert\Bar{\rho}_i\rVert,\ \forall n_i\in\mathcal{N}_i,
    \end{align}
    where $\epsilon\in(0,1)$, $\eta_i=[\eta_{i,1},\ldots,\eta_{i,n}]$, $\eta_{i,j}=B_{f_{i,j}}+\sigma\sqrt{2(\gamma_{i,j}+1+ln(\frac{1}{\epsilon_{hn}}))}$, $\forall j\in\{1,\ldots,n\}$, $B_{i,j}\geq\lVert f_{i,j}\rVert_{k_{i,j}}$, $\epsilon_{nh}=\frac{\epsilon}{n h}$, $h$ is the number of subsystems in (\ref{equ:sys}), $\Bar{\rho}_i:=\max_{n_i\in\mathcal{N}_i}\rho_i(n_i)$, and $\gamma_{i,j}$ is the maximum information gain (refer remark \ref{remark:infoGain}).
\end{lemma}
\begin{proof}
This proof is inspired by \cite[Lemma 2]{GPBackStepping}. For every $i$, $f_i$ is an $n$-dimensional function, where $f_{i,j}:\mathcal{N}_i\rightarrow\R$, $j\in\{1,\ldots,n\}$ is a scalar function. Hence, from \cite[Theorem 2]{bounds}, we have the following that holds with probability $1-\epsilon_{nh}$,
    \begin{align*}
        |f_{i,j}(n_i)-\mu_{i,j}(n_i)|&\leq\eta_{i,j}\rho(n_i),\forall n_i\in\mathcal{N}_i,\\
        \lVert f_{i,j}(n_i)-\mu_{i,j}(n_i)\rVert^2&\leq\eta^2_{i,j}\rho^2_{i,j}(n_i),\forall n_i\in\mathcal{N}_i.
    \end{align*}
    This implies that $$\mathbb{P}\left\{\lVert f_{i,j}(n_i)-\mu_{i,j}(n_i)\rVert^2>\eta^2_{i,j}\rho^2_{i,j}(n_i),\forall n_i\in\mathcal{N}_i\right\}<\epsilon_{nh}.$$ By applying union bounds,
    \begin{align*}
        \mathbb{P}\left\{\bigcup_{j=1}^n\lVert f_{i,j}(n_i)-\mu_{i,j}(n_i)\rVert^2>\eta^2_{i,j}\rho^2_{i,j}(n_i),\forall n_i\in\mathcal{N}_i\right\}\nonumber<n\epsilon_{nh}\nonumber,\\
        \mathbb{P}\left\{\bigcap_{j=1}^n\lVert f_{i,j}(n_i)-\mu_{i,j}(n_i)\rVert^2\leq\eta^2_{i,j}\rho^2_{i,j}(n_i),\forall n_i\in\mathcal{N}_i\right\}\nonumber\geq1-\epsilon_{h},
    \end{align*}
    where $\epsilon_h=n\epsilon_{nh}$.
    Now, we can say that the following holds with a probability of at least $1-\epsilon_h$ and $\forall n_i\in\mathcal{N}_i$,
    \begin{align}\label{equ:ineqI}
        &\lVert f_i(n_i)-\mu_i(n_i)\rVert^2\leq\lVert\eta_i^{\top}\rho_i(n_i)\rVert^2\nonumber\\
        \implies&\lVert f_i(n_i)-\mu_i(n_i)\rVert\leq\lVert\eta_i\rVert\lVert\rho_i(n_i)\rVert.
    \end{align}
    We use Cauchy-Schwartz inequality here. Since $\Bar{\rho}_i=\max_{n_i\in\mathcal{N}_i}\rho_i(n_i)$ exists, similar to \cite{JagtapGP}, we can rewrite (\ref{equ:ineqI}) as $\mathbb{P}\left\{\lVert f_i(n_i)-\mu_i(n_i)\rVert\in\{d|d\in[0,\lVert\eta_i\rVert\lVert\Bar{\rho}_i\rVert]\},\forall n_i\in\mathcal{N}_i\right\}\geq1-\epsilon_h$. Hence, $\mathbb{P}\left\{\lVert f_i(n_i)-\mu_i(n_i)\rVert\leq\lVert\eta_i\rVert\lVert\Bar{\rho}_i\rVert,\forall n_i\in\mathcal{N}_i\right\}\geq1-\epsilon_h$. Again, by applying union bounds over $i$ and given $\mathbb{P}\left\{\lVert f_i(n_i)-\mu_i(n_i)\rVert>\lVert\eta_i\rVert\lVert\Bar{\rho}_i\rVert,\forall n_i\in\mathcal{N}_i\right\}<1-\epsilon_h$,
    \begin{align*}
        \mathbb{P}\left\{\bigcup_{i=1}^h\lVert f_i(n_i)-\mu_i(n_i)\rVert>\lVert\eta_i\rVert\lVert\Bar{\rho}_i\rVert,\forall n_i\in\mathcal{N}_i\right\}<h\epsilon_h,\\
        \implies\mathbb{P}\left\{\bigcap_{i=1}^h\lVert f_i(n_i)-\mu_i(n_i)\rVert\leq\lVert\eta_i\rVert\lVert\Bar{\rho}_i\rVert,\forall n_i\in\mathcal{N}_i\right\}\geq1-\epsilon.
    \end{align*}
    This ends the proof.
\end{proof}
If $B_{f_i}$ is not available, a guess and doubling strategy can be employed to obtain an estimate \cite{InfoGain}. 
\begin{remark}\label{remark:infoGain}
        The information gain $\gamma_{i,j}$ can be defined as in \cite{GPBackStepping}, and it is a measure of the reduction of uncertainty achievable given that the measurements are taken in the best possible conditions. $\gamma_{i,j}$ has sub-linear dependence on $N$, which is the size of the dataset $\mathcal{D}_i$ for most kernels in use. Hence, with an increase in size of $\mathcal{D}_i$, the model error decreases. For further discussion, the readers are referred to \cite{InfoGain}.
\end{remark}
It is also possible to provide a deterministic bound on the model error if the RKHS norm's bound, $\lVert f_{i,j}\rVert_{k_{i,j}}\leq B_{i,j}$, can be computed. Even though this computation is hard, it can be done with a Lipschitz-like assumption on $f$, as shown in the following lemma.
    \begin{lemma}{\cite[Lemma 1]{adnaneDeterministic}}\label{lemma:B_idef}
        Given a kernel function $k_{i,j}$ and a function $f_{i,j}$ such that $|f_{i,j}(n_i)-f_{i,j}(n'_i)|\leq\mathcal{L}_{i,j}\sqrt{\lVert n_i-n'_i\rVert}$, for all $n_i,n'_i\in\mathcal{N}_i$, where $L_{i,j}\in\R^+$, $B_{i,j}=\frac{\mathcal{L}_{i,j}}{\sqrt{2\lVert\frac{\partial k_{i,j}}{\partial n_i}\rVert}}$.
    \end{lemma}
    Using this result, we can now compute the deterministic bound as presented in the following lemma.
    \begin{lemma}{\cite[Lemma 2]{adnaneDeterministic}}\label{lemma:deterministicBound}
        Given the system (\ref{equ:sys}) with assumptions \ref{assume:system}, \ref{assume:RKHS}, and \ref{assume:availabledata} and GP approximation with mean $\mu_i(n_i)$ and standard deviation $\rho_i(n_i)$ as given in (\ref{equ:mean}) and (\ref{equ:variance}), respectively, the following holds $\forall n_i\in\mathcal{N}_i$ with probability $1$,
        \begin{align}\label{equ:deterministicBound}
            \lVert f_i(n_i)-\mu_i(n_i)\rVert\leq\lVert\Tilde{\eta}_i\rVert\lVert\Bar{\rho}_i\rVert,
        \end{align} where $\Tilde{\eta}_i=\left[\Tilde{\eta}_{i,1},\ldots,\Tilde{\eta}_{i,n}\right]$, $\Tilde{\eta}_{i,j}=\sqrt{B_{i,j}^2-y_{i,j}^{\top}(K_{i,j}+\sigma^2\mathbf{I}_N)^{-1}y_{i,j}+N}$, $B_{i,j}$ is as defined in Lemma \ref{lemma:B_idef}, $y_{i,j}$ and $K_{i,j}$ are as defined in (\ref{equ:mean_i}) and (\ref{equ:variance_i}), respectively, and $N$ is the number of data points.
    \end{lemma}
    The deterministic bound (\ref{equ:deterministicBound}) is conservative, and this is evident in the case studies. Using the learned dynamics along with the probabilistic and deterministic bounds on model errors provided in lemmas \ref{lemma:probBound} and \ref{lemma:deterministicBound}, respectively, we now proceed to formulate stabilization controllers using the backstepping scheme presented in the next section.
\section{Backstepping Control Design Scheme}\label{Sec:control design}
With the learned model and the bounds on model error presented in the previous section, we now provide a backstepping control design scheme that synthesizes controllers for enforcing $\delta$-ISpS properties on the system (\ref{equ:sys}). The main result of the paper on the backstepping control design scheme is presented in the following theorem.
\begin{theorem}\label{theorem:SFControl}
    Given a control system $\Sigma=(\mathcal{X},U,\mathcal{U},f)$ of the form (\ref{equ:sys}) satisfying assumptions \ref{assume:system}, \ref{assume:RKHS}, and \ref{assume:availabledata} and corresponding approximated GP model with mean and standard deviation  given by (\ref{equ:mean}) and (\ref{equ:variance}), respectively; the following state feedback control law:
    \begin{align}\label{equ:controlaw}
        \upsilon=&\frac{1}{b_h}\Big(-\mu_h(x)-b_{h-1}\left(\xi_{h-1}-\psi_{h-2}\right)-\lambda_h\left(\xi_h-\psi_{h-1}\right)+\sum_{i=1}^{h-1}\frac{\partial \psi_{h-1}}{\partial \xi_i}\left(\mu_i+b_i\xi_{i+1}\right)+\hat{\upsilon}\Big),
    \end{align}
    where 
    \begin{align}
        \psi_i=&-\frac{1}{b_i}\Big(\mu_i+b_{i-1}(\xi_{i-1}-\psi_{i-2})+\lambda_i(\xi_i-\psi_{i-1})-\sum_{j=1}^{i-1}\frac{\partial \psi_{i-1}}{\partial \xi_j}(\mu_j+b_j\xi_{j+1})\Big),
    \end{align}
    for all $i\in\{1,\ldots,h-1\}$, $\psi_{-1}=\psi_{0}=b_0=\xi_0=0$, $\lambda_1>1$, $\lambda_i>1+\sum_{j=1}^{i-1}{L}_{\psi_{i-1}\xi_j}$ for all $i\in\{2,\ldots,h-1\}$, $\lambda_h>1.5+\sum_{j=1}^{h-1}{L}_{\psi_{h-1}\xi_j}$ and ${L}_{\psi_j\xi_k}\geq\frac{\partial \psi_j}{\partial x_k}$, renders the controlled system $\delta$-ISpS with respect to the input $\hat{\upsilon}$ with a probability of at least $1-\epsilon$.
\end{theorem}
\begin{proof}
    To show that the controlled system is $\delta$-ISpS with respect to input as shown in Definition \ref{def:deltaISpS}, we have to prove that it admits a $\delta$-ISpS Lyapunov function as defined in Definition \ref{def:deltaISpSLyapunov}.

    First, consider the following transformation,
    \begin{align}
        \zeta=\varphi(\xi)=\begin{bmatrix}
            \zeta_1\\
            \zeta_2\\
            \vdots\\
            \zeta_k\\
            \vdots\\
            \zeta_h
        \end{bmatrix}=\begin{bmatrix}
            \xi_1\\
            \xi_2-\psi_1\\
            \vdots\\
            \xi_k-\psi_{k-1}\\
            \vdots\\
            \xi_h-\psi_{h-1}
        \end{bmatrix},
    \end{align}
    where $ \psi_i=-\frac{1}{b_i}\big(\mu_i+b_{i-1}(\xi_{i-1}-\psi_{i-2})+\lambda_i(\xi_i-\psi_{i-1})-\sum_{j=1}^{i-1}\frac{\partial \psi_{i-1}}{\partial \xi_j}(\mu_j+b_j\xi_{j+1})\big)$, for $\lambda_i>0$, for all $i\in\{1,\ldots,h-1\}$, $\zeta\in\mathcal{Z}$, $\mathcal{Z}=\prod_{i=1}^h\mathcal{Z}_i$ and $\prod_{i=1}^h\mathcal{Z}_i=\varphi(\prod_{i=1}^h\mathcal{X}_i)$. Also note that we use $f_i$ and $\mu_i$ to represent $f_i(\nu_i)$ and $\mu_i(\nu_i)$, respectively. The same applies to the transformed space. By applying the transformation and the control input, the system (\ref{equ:sys}) can be written as:
    \begin{equation}
        \begin{split}
        \dot{\zeta_1} &= f_1-\mu_1+b_1\zeta_2-\lambda_1\zeta_1,\nonumber\\
        \dot{\zeta_k} &= f_k-\mu_k+b_k\zeta_{k+1}-b_{k-1}\zeta_{k-1}-\lambda_k\zeta_k\nonumber\\
        &-\sum_{j=1}^{k-1}\frac{\partial \psi_{k-1}}{\partial(\zeta_j+\psi_{j-1})}(f_j-\mu_j), k\in\{2,\ldots,h-1\},\nonumber\\
        \dot{\zeta_h} &= f_h-\mu_h-b_{h-1}\zeta_{h-1}-\lambda_h\zeta_h\nonumber-\sum_{j=1}^{h-1}\frac{\partial \psi_{h-1}}{\partial(\zeta_j+\psi_{j-1})}(f_j-\mu_j)+\hat{\upsilon}.
        \end{split}
    \end{equation}
    We first define a candidate Lyapunov function for the $\zeta_1$-subsystem, for all $z_1,z'_1\in\mathcal{Z}_1$, as $V_1(z_1,z'_1)=(z_1-z'_1)^{\top}(z_1-z'_1)$. Now, taking the time derivative, one obtains:  
    \begin{align}\label{equ:V1ineq}      &\dot{V}_1(z_1,z'_1)=2(z_1-z'_1)^{\top}\dot{z}_1-2(z_1-z'_1)^{\top}\dot{z}'_1\nonumber\\
        &=2(z_1-z'_1)^{\top}(f_1-\mu_1+b_1z_2-\lambda_1z_1)\nonumber-2(z_1-z'_1)^{\top}(f'_1-\mu'_1+b_1z'_2-\lambda_1z'_1)\nonumber\\
        &=-2\lambda_1\| z_1-z'_1\|^2+2b_1(z_1-z'_1)^{\top}(z_2-z'_2)\nonumber+2(z_1-z'_1)^{\top}(f_1-\mu_1)-2(z_1-z'_1)^{\top}(f'_1-\mu'_1)\nonumber\\
        &\leq-2\lambda_1\| z_1-z'_1\|^2+2b_1\| z_1-z'_1\|\|z_2-z'_2\|+4\| z_1-z'_1\|\|f_1-\mu_1\|\nonumber\\
        &\leq-(2\lambda_1-b_1-2)\| z_1-z'_1\|^2+b_1\| z_2-z'_2\|^2+2\| \eta_1\|^2\|\Bar{\rho}_1\|^2.
    \end{align}
    The inequalities appear due to the implementation of Cauchy-Schwartz and Young's inequalities.
    Now, we consider a candidate Lyapunov function for the subsystem $\zeta_k$, where for every $z_k,z'_k\in\mathcal{Z}_k$ and any $k\in\{2,\ldots,h-2\}$, $V_k(z_k,z'_k)=(z_k-z'_k)^{\top}(z_k-z'_k)$. Now,
    {\allowdisplaybreaks
\begin{align}
    &\dot V_k(z_k,z'_k)=2(z_k-z'_k)^{\top}\dot z_k-2(z_k-z'_k)^{\top}\dot z'_k\nonumber\\
    &=2(z_k-z'_k)^{\top}\Big(f_k-\mu_k+b_kz_{k+1}-b_{k-1}z_{k-1}-\lambda_kz_k-\sum_{j=1}^{k-1}\frac{\partial \psi_{k-1}}{\partial(\zeta_j+\psi_{j-1})}(f_j-\mu_j)\Big)\nonumber\\
    &-2(z_k-z'_k)^{\top}\Big(f'_k-\mu'_k\nonumber+b_kz'_{k+1}-b_{k-1}z'_{k-1}-\lambda_kz'_k\nonumber-\sum_{j=1}^{k-1}\frac{\partial \psi_{k-1}}{\partial(\zeta_j+\psi_{j-1})}(f_j'-\mu_j')\Big)\nonumber\\
    &=2(z_k-z'_k)^{\top}(f_k-\mu_k)+2b_k(z_k-z'_k)^{\top}(z_{k+1}-z'_{k+1})\nonumber\\
    &-2b_{k-1}(z_k-z'_k)^{\top}(z_{k-1}-z'_{k-1})-2\lambda_k\lVert z_k-z'_k\rVert^2\nonumber\\
    &+2(z_k-z'_k)^{\top}\Big(\sum_{j=1}^{k-1}\frac{\partial \psi_{k-1}}{\partial(\zeta_j+\psi_{j-1})}(\mu_j-f_j)\nonumber+\frac{\partial \psi'_{k-1}}{\partial(\zeta'_j+\psi'_{j-1})}(f'_j-\mu'_j)\Big)-2(z_k-z'_k)^{\top}(f'_k-\mu'_k)\nonumber\\
    &\leq2(z_k-z'_k)^{\top}(f_k-\mu_k)+2b_k(z_k-z'_k)^{\top}(z_{k+1}-z'_{k+1})\nonumber-2b_{k-1}(z_k-z'_k)^{\top}(z_{k-1}-z'_{k-1})-2\lambda_k\lVert z_k-z'_k\rVert^2\nonumber\\
    &-2\sum_{j=1}^{k-1}\mathcal{L}_{\psi_{k-1}\xi_j}(z_k-z'_k)^{\top}((f_j-\mu_j)-(f'_j-\mu'_j))\nonumber-2(z_k-z'_k)^{\top}(f'_k-\mu'_k)\nonumber\\
    &\leq2\lVert z_k-z'_k\rVert\lVert f_k-\mu_k\rVert+2b_k\lVert z_k-z'_k\rVert\lVert z_{k+1}-z'_{k+1}\rVert\nonumber-2b_{k-1}\lVert z_k-z'_k\rVert\lVert z_{k-1}-z'_{k-1}\rVert-2\lambda_k\lVert z_k-z'_k\rVert^2\nonumber\\
    &-2\sum_{j=1}^{k-1}{L}_{\psi_{k-1}\xi_j}\lVert z_k-z'_k\rVert(\lVert f_j-\mu_j\rVert-\lVert f_j'-\mu_j'\rVert)\nonumber-2\lVert z_k-z'_k\rVert\lVert f'_k-\mu'_k\rVert\nonumber\\
    &\leq4\lVert z_k-z'_k\rVert\lVert \eta_k\rVert\lVert \Bar{\rho}_k\rVert+2b_k\lVert z_k-z'_k\rVert\lVert z_{k+1}-z'_{k+1}\rVert\nonumber-2b_{k-1}\lVert z_k-z'_k\rVert\lVert z_{k-1}-z'_{k-1}\rVert-2\lambda_k\lVert z_k-z'_k\rVert^2\nonumber\\
    &+4\sum_{j=1}^{k-1}{L}_{\psi_{k-1}\xi_j}\lVert z_k-z'_k\rVert\lVert \eta_j\rVert\lVert \Bar{\rho}_j\rVert\nonumber\\
    &\leq-(2\lambda_k-2-b_k+b_{k-1}-2\sum_{j=1}^{k-1}{L}_{\psi_{k-1}\xi_j})\lVert z_k-z'_k\rVert^2\nonumber+b_k\lVert z_{k+1}- z'_{k+1}\rVert^2-b_{k-1}\lVert z_{k-1}-z'_{k-1}\rVert^2\nonumber\\
    &+2\sum_{j=1}^{k-1}{L}_{\psi_{k-1}\xi_j}\lVert \eta_j\rVert^2\lVert \Bar{\rho}_j\rVert^2+2\lVert \eta_k\rVert^2\lVert\Bar{\rho}_k\rVert^2.
    \label{equ:Vkineq}
    \end{align}
    }
    For the $\zeta_{k+1}$-subsystem, we define the candidate Lyapunov function for any $z_{k+1},z'_{k+1}\in\mathcal{Z}_{k+1}$,
        $V_{k+1}(z_{k+1},z'_{k+1})=(z_{k+1}-z'_{k+1})^{\top}(z_{k+1}-z'_{k+1})$. Similar to (\ref{equ:Vkineq}), the following holds,
\allowdisplaybreaks{
\begin{align}\label{equ:Vk+1ineq}
        &\dot V_{k+1}(z_{k+1},z'_{k+1})\nonumber\leq-(2\lambda_{k+1}-2-b_{k+1}+b_{k}-2\sum_{j=1}^{k}\mathcal{L}_{\psi_{k}\xi_j})\lVert z_{k+1}-z'_{k+1}\rVert^2\nonumber\\&\ \ \ \ +b_{k+1}\lVert z_{k+2}-x'_{k+2}\rVert^2-b_{k}\lVert z_{k}-z'_{k}\rVert^2 +2\sum_{j=1}^{k}\mathcal{L}_{\psi_{k}\xi_j}\lVert \eta_j\rVert^2\lVert \Bar{\rho}_j\rVert^2+2\lVert \eta_{k+1}\rVert^2\lVert\Bar{\rho}_{k+1}\rVert^2.
    \end{align}
    Finally, for the $\zeta_h$-subsystem, the candidate Lyapunov function for any $z_h,z'_h\in\mathcal{Z}_h$ is given by $V_h(z_h,z'_h)=(z_h-z'_h)^{\top}(z_h-z'_h)$ and
\begin{align}\label{equ:Vhineq}
        &\dot V_h(z_h,z'_h)=2(z_h-z'_h)^{\top}\dot z_h-2(z_h-z'_h)^{\top}\dot z'_h
        =2(z_h-z'_h)^{\top}\Big(f_h-\mu_h-b_{h-1}z_{h-1}-\lambda_hz_h\nonumber\\
        &-\sum_{j=1}^{h-1}\frac{\partial\psi_{h-1}}{\partial(z_j+\psi_{j-1})}[f_j-\mu_j]+\hat{u}\Big)\nonumber-2(z_h-z'_h)^{\top}\Big(f'_h-\mu'_h-b_{h-1}z'_{h-1}-\lambda_hz'_h\nonumber\\
        &-\sum_{j=1}^{h-1}\frac{\partial\psi'_{h-1}}{\partial(z'_j+\psi'_{j-1})}(f'_j-\mu'_j)+\hat{u}'\Big)\nonumber\\
        &=2(z_h-z'_h)^{\top}(f_h-\mu_h)-2b_{h-1}(z_h-z'_h)^{\top}(z_{h-1}-z'_{h-1})\nonumber\\
        &-2\lambda_h\lVert z_h-z'_h\rVert^2+2(z_h-z'_h)^{\top}\sum_{j=1}^{h-1}\frac{\partial\psi_{h-1}}{\partial(z_j+\psi_{j-1})}[\mu_j-f_j]\nonumber\\
        &+2(z_h-z'_h)^{\top}\sum_{j=1}^{h-1}\frac{\partial\psi'_{h-1}}{\partial(z'_j+\psi'_{j-1})}(f'_j-\mu'_j)\nonumber-2(z_h-z'_h)^{\top}(f'_h-\mu'_h)+2(z_h-z'_h)^{\top}(\hat{u}-\hat{u}')\nonumber\\
        &\leq2\lVert z_h-z'_h\rVert\lVert f_h-\mu_h\rVert-2b_{h-1}\lVert z_h-z'_h\rVert\lVert z_{h-1}-z'_{h-1}\rVert\nonumber\\
        &-2\lambda_h\lVert z_h-z'_h\rVert^2-2\sum_{j=1}^{h-1}\mathcal{L}_{\psi_{h-1}\xi_j}\lVert z_h-z'_h\rVert\lVert f_j-\mu_j\rVert\nonumber+2\sum_{j=1}^{h-1}\mathcal{L}_{\psi_{h-1}\xi_j}\lVert z_h-z'_h\rVert\lVert f'_j-\mu'_j\rVert\nonumber\\
        &-2\lVert z_h-z'_h\rVert\lVert f'_h-\mu'_h\rVert+2\lVert z_h-z'_h\rVert\lVert \hat{u}-\hat{u}'\rVert\nonumber\\
        &\leq-2\lambda_h\lVert z_h-z'_h\rVert^2-2b_{h-1}\lVert z_h-z'_h\rVert\lVert z_{h-1}-z'_{h-1}\rVert\nonumber+4\lVert z_h-z'_h\rVert\rVert\eta_h\rVert\lVert\Bar{\rho}_h\rVert\nonumber\\
        &+4\sum_{j=1}^{k-1}\mathcal{L}_{\psi_{h-1}\xi_j}\lVert z_h-z'_h\rVert\lVert\eta_j\rVert\lVert\Bar{\rho}_j\rVert\nonumber+2\lVert z_h-z'_h\rVert\lVert \hat{u}-\hat{u}'\rVert\nonumber\\
        &\leq-(2\lambda_h+b_{h-1}-3-2\sum_{j=1}^{h-1}\mathcal{L}_{\psi_{h-1}\xi_j})\lVert z_h-z'_h\rVert^2\nonumber\\
        &-b_{h-1}\lVert z_{h-1}-z'_{h-1}\rVert+2\sum_{j=1}^{h-1}\mathcal{L}_{\psi_{h-1}\xi_j}\lVert\eta_j\rVert^2\lVert\Bar{\rho}_j\rVert^2+2\lVert\eta_h\rVert^2\lVert\Bar{\rho}_h\rVert^2+\lVert\hat{u}-\hat{u}'\rVert^2.
    \end{align}
    }
    Let the Lyapunov function of the system $\Sigma$ given in (\ref{equ:sys}) be,
    \begin{align*}       V(z,z')=\sum_{k=1}^hV_k(z_k,z'_k),
    \end{align*}for any $z,z'\in\mathcal{Z}$, $z=[z_1^{\top},\ldots,z_h^{\top}]^{\top}$ and $z'=[z_1^{'\top},\ldots,z_h^{'\top}]^{\top}$. From (\ref{equ:V1ineq})-(\ref{equ:Vhineq}),
\begin{align}
&\dot V\leq\hspace{-0.1em}-(2\lambda_1\hspace{-0.1em}-\hspace{-0.1em}b_1\hspace{-0.1em}-\hspace{-0.1em}2)\lVert z_1\hspace{-0.1em}-\hspace{-0.1em}z'_1\rVert^2\hspace{-0.15em}+\hspace{-0.15em}b_1\lVert z_2\hspace{-0.1em}-\hspace{-0.1em}z'_2\rVert^2\hspace{-0.15em}+\hspace{-0.15em}2\lVert\eta_1\rVert^2\lVert\Bar{\rho}_1\rVert^2\nonumber\\
&\hspace{-0.1em}+\hspace{-0.1em}\sum_{k=2}^{h-1}\hspace{-0.2em}\big(\hspace{-0.2em}-(2\lambda_k-2-b_k+b_{k-1}-2\sum_{j=1}^{k-1}\mathcal{L}_{\psi_{k-1}\xi_j})\lVert z_k-z'_k\rVert^2\nonumber\\
&+b_k\lVert z_{k+1}-z'_{k+1}\rVert^2-b_{k-1}\lVert z_{k-1}-z'_{k-1}\rVert^2\nonumber+2\sum_{j=1}^{k-1}\mathcal{L}_{\psi_{k-1}\xi_j}\lVert\eta_j\rVert^2\lVert\Bar{\rho}_j\rVert^2+2\lVert\eta_k\rVert^2\lVert\Bar{\rho}_k\rVert^2\big)\nonumber\\
&+\big(-(2\lambda_h+b_{h-1}-3-2\sum_{j=1}^{h-1}\mathcal{L}_{\psi_{h-1}\xi_j})\lVert z_h-z'_h\rVert^2\nonumber\\
&-b_{h-1}\lVert z_{h-1}-z'_{h-1}\rVert^2+2\sum_{j=1}^{h-1}\mathcal{L}_{\psi_{h-1}\xi_j}\lVert\eta_j\rVert^2\lVert\Bar{\rho}_j\rVert^2\nonumber+2\lVert\eta_h\rVert^2\lVert\Bar{\rho}_h\rVert^2+\lVert\hat{u}-\hat{u}'\rVert^2\big),\nonumber\\
&\leq-(2\lambda_1-2)\lVert z_1-z'_1\rVert^2-\sum_{k=2}^{h-1}(2\lambda_k-2-2\sum_{j=1}^{k-1}\mathcal{L}_{\psi_{k-1}\xi_j})\nonumber\lVert z_k-z'_k\rVert^2\\
&-(2\lambda_h-3-2\sum_{j=1}^{h-1}\mathcal{L}_{\psi_{h-1}\xi_j})\lVert z_h-z'_h\rVert^2\nonumber\\
&+\lVert\hat{u}-\hat{u}'\rVert^2+2\sum_{k=1}^h\lVert\eta_k\rVert^2\lVert\Bar{\rho}_k\rVert^2\nonumber+2\sum_{k=2}^h\sum_{j=1}^{k-1}\mathcal{L}_{\psi_{k-1}\xi_j}\lVert\eta_j\rVert^2\lVert\Bar{\rho}_j\rVert^2,\nonumber\\
&\leq-\sum_{k=1}^{h}k_k\lVert z_k-z'_k\rVert^2+\lVert\hat{u}-\hat{u}'\rVert^2+2\sum_{k=1}^h\lVert\eta_k\rVert^2\lVert\Bar{\rho}_k\rVert^2+2\sum_{k=2}^h\sum_{j=1}^{k-1}\mathcal{L}_{\psi_{k-1}\xi_j}\lVert\eta_j\rVert^2\lVert\Bar{\rho}_j\rVert^2.
\end{align}
By defining $\lambda_1>1$, $\lambda_k>1+\sum_{j=1}^{k-1}\mathcal{L}_{\psi_{k-1}\xi_j}$ for all $k\in\{1,\ldots,h-1\}$, $\lambda_h>1.5+\sum_{j=1}^{h-1}\mathcal{L}_{\psi_{h-1}\xi_j}$, $k=\min(k_1,k_2,\ldots,k_h)$ and $\Tilde{c}=2\sum_{k=1}^h\lVert\eta_k\rVert^2\lVert\Bar{\rho}_k\rVert^2+2\sum_{k=2}^h\sum_{j=1}^{k-1}\mathcal{L}_{\psi_{k-1}\xi_j}\lVert\eta_j\rVert^2\lVert\Bar{\rho}_j\rVert^2$, we get   \begin{align}\label{equ:finalineq}
        \dot V(z,z')\leq -kV(z,z')+\sigma(\lVert\hat{u}-\hat{u}'\rVert)+\Tilde{c}.
    \end{align}
Since $k_k>0$, $\forall d\in\{1,\ldots,h\}$, (\ref{equ:finalineq}) satisfies the condition $(ii)$ in definition (\ref{def:deltaISpSLyapunov}). In addition, by defining $\mathbf{d}$ as the natural Euclidean norm, $\underline{\alpha}(r)=r$ and $\overline{\alpha}(r)=2r^2$, it is easy to show that the Lyapunov function also satisfies condition $(i)$. Therefore, from Theorem \ref{theorem:ISpSLyapunov}, we can easily say that
\begin{align}
\lVert \varphi(\xi_{x\upsilon}(t))-\varphi(\xi_{x'\upsilon'}(t))\rVert&\leq\beta(\lVert \varphi(x)-\varphi(x)'\rVert,t)+\gamma(\lVert\upsilon-\upsilon'\rVert_\infty)+c,
    \end{align}where
    \begin{align*}  &\beta(r,s)=\underline{\alpha}^{-1}(3e^{-ks}\overline{\alpha}(r))
        =6e^{-ks}r^2,\\    &\gamma(r)=\underline{\alpha}^{-1}\left(\frac{3}{ek}\sigma(r)\right)=\frac{3}{ek}\sigma(r), \forall r,s\in\R^+_0\\ &c=\underline{\alpha}^{-1}\left(\frac{3\Tilde{c}}{ek}\right)=\frac{3\Tilde{c}}{ek}
    \end{align*}
    are the class-$\mathcal{KL}$, class-$\mathcal{K}_\infty$ functions and constant, respectively. Defining metric $\mathbf{d}(x,x')=\lVert\varphi(x)-\varphi(x')\rVert$,
    \begin{align*}
        \mathbf{d}(\xi_{x\hat{v}}(t),\xi_{x'\hat{v}'}(t))\leq\beta(\mathbf{d}(x,x'),t)+\gamma(\lVert \hat{v}-\hat{v}'\rVert_\infty)+c.
    \end{align*}
    This proves that the control law (\ref{equ:controlaw}) renders the control system $\Sigma$ given by (\ref{equ:sys}) $\delta$-ISpS.
\end{proof}
\begin{remark}
    Note that we assume that the term $\frac{\partial \psi_j}{\partial x_k}$ is bounded by $\mathcal{L}_{\psi_jx_k}$ for every $j\in\{1,\ldots,h-1\}$ and $k\in\{1,\ldots,j\}$. This is a valid assumption since $\psi_i$ is a combination of the mean functions of the trained $GP$s, linear combinations of the states and defined in a compact state-space $\mathcal{X}$.
\end{remark}
\begin{remark}\label{remark:probofISpS}
    Since the model error between the approximation and the actual function is bounded probabilistically, the system is $\delta$-ISpS with that same high probability $(1-\epsilon)$.
\end{remark}
\begin{remark}
    Since we have a value for $\Tilde{c}$, a non-disappearing disturbance that produces a mismatch in trajectory even after an arbitrarily long time can be quantized to be ${c}=\underline{\alpha}^{-1}\big(\frac{6}{ek}[\sum_{k=1}^h\lVert\eta_k\rVert^2\lVert\Bar{\rho}_k\rVert^2$ +$\sum_{k=2}^h\sum_{j=1}^{k-1}L_{\psi_{k-1}\xi_j}\lVert\eta_j\rVert^2\lVert\Bar{\rho}_j\rVert^2]\big)$, where $\underline{\alpha}\in\mathcal{K}_\infty$ is the identity function. This means that even at an arbitrarily large value of $t$, the trajectories of the system might not exactly converge to each other but might differ with a value $c$.
\end{remark}
\begin{corollary}
    Given the system (\ref{equ:sys}) approximated by mean $\mu(x)$ and standard deviation $\rho(x)$ as defined in (\ref{equ:mean}) and (\ref{equ:variance}), respectively, $\Bar{\rho}=\max_{x\in\mathcal{X}}\rho(x)$ and control law as shown in (\ref{equ:controlaw}), if the model error is bounded as shown in Lemma \ref{lemma:deterministicBound}, then (\ref{equ:controlaw}) renders the system $\delta$-ISpS with respect to $\hat{\upsilon}$ with a probability of $1$.
\end{corollary}
\begin{proof}
    The corollary is a direct consequence of Theorem \ref{theorem:SFControl} and Lemma \ref{lemma:deterministicBound}.
\end{proof}
The only change here, due to the more conservative bound on the model error, is in the value of ${c}$, which is larger. This means that for a larger non-disappearing perturbation in the upper bound in (\ref{equ:deltaISpS}), the system is $\delta$-ISpS with probability $1$, and the trajectories never diverge more than that bound.
\section{Case Study}
This paper considers two case studies, $(i)$ a magnetic levitation system and $(ii)$ a two-link manipulator. We first explain the experiment run on magnetic levitation and show the corresponding results, followed by the case study of two link manipulators.
\subsection{Magnetic Levitation System}
The magnetic levitation system \cite{jeltsema2009multidomain} represented as
\begin{align}\label{equ:magnetic_lev}
    \dot \xi_1&=\frac{\xi_2}{M},\nonumber\\
    \dot \xi_2 &= \frac{\xi_3}{2\alpha}-Mg,\nonumber\\
    \dot \xi_3&=\frac{-2R}{\alpha}(1-\xi_1)\xi_3+2\sqrt{\xi_3}v.
\end{align}
where $\xi_1$ represents the displacement of the ball, $\xi_2$ signifies the momentum associated with it, $\xi_3$ represents the square of flux linkage in the electromagnetic coil, and $v$ is the voltage applied across the electromagnetic coil. $M$, $g$, and $R$ are the ball's mass, acceleration due to gravity, and coil resistance, respectively. $\alpha>0$ is a constant that depends on the number of turns in the electromagnetic coil. We consider a compact set $\mathcal{X}=[0,4]\times[-6,6]\times[0,18]$. It is obvious that $f$ is continuous and has a finite RKHS norm satisfying Assumption \ref{assume:RKHS}.

We learn the unknown model using the Gaussian process with $200$ data samples of $x$ and $y=f(x)+w$, where $w\sim\mathcal{N}(0,\rho_f^2\mathbf{I}_2)$ and $\rho_f=0.01$, collected by simulating the system with several initial states chosen randomly. The kernel used is a squared-exponential kernel \cite{GPBook} given by: $k(x,x')=\rho_k^2\exp\left(\sum_{i=1}^2\frac{(x_i-x'_i)^2}{-2l_i^2}\right)$, where $\rho_k=119$ is the signal variance and $l_1=6$ and $l_2=1.45\times10^{4}$, and $l_3=14.3$ are the length scales. These parameters are obtained using the Limited memory Broyden-Fletcher-Goldfarb-Shanno (L-BFGS-B) algorithm \cite{ACMOpti}. The inferred mean and variance are as defined in (\ref{equ:mean}) and (\ref{equ:variance}) respectively with $\Bar{\rho} = 0.000346$. Since the computation of $\lVert f\rVert_{k}$ and $\gamma$ is a hard problem, we obtain the probability bound on the accuracy of the learned model using the Monte-Carlo approach. For a preset value of $\lVert\eta\rVert\lVert\Bar{\rho}\rVert=0.00188$, we obtained an interval for the probability in (\ref{equ:deterministicBound}) such that $(1-\epsilon)^n\in\left[0.984,0.986\right]$ with a confidence of $1-10^{-10}$ using $10^6$ realizations. Thus, following Remark \ref{remark:probofISpS}, we can say that the controlled system with the controller designed as shown in Section \ref{Sec:control design} is $\delta$-ISpS with a probability of at least $0.984$ if ${c}=7.985\times10^{-6}$.

In addition, we also compare the value of $\lVert\Tilde{\eta}\rVert\lVert\Bar{\rho}\rVert = 0.0049$ for $\lVert\Tilde{\eta}\rVert=34.964$ computed as in Lemma \ref{lemma:deterministicBound} with the value of $\lVert\eta\rVert\lVert\Bar{\rho}\rVert=0.02$ computed by the Monte-Carlo approach such that the probability interval is $[1.0,1.0]$. The conservatism of the bound according to Lemma \ref{lemma:deterministicBound} is visible in this comparison.

\begin{figure}\label{fig:levTraj}
\includegraphics[width=\linewidth]{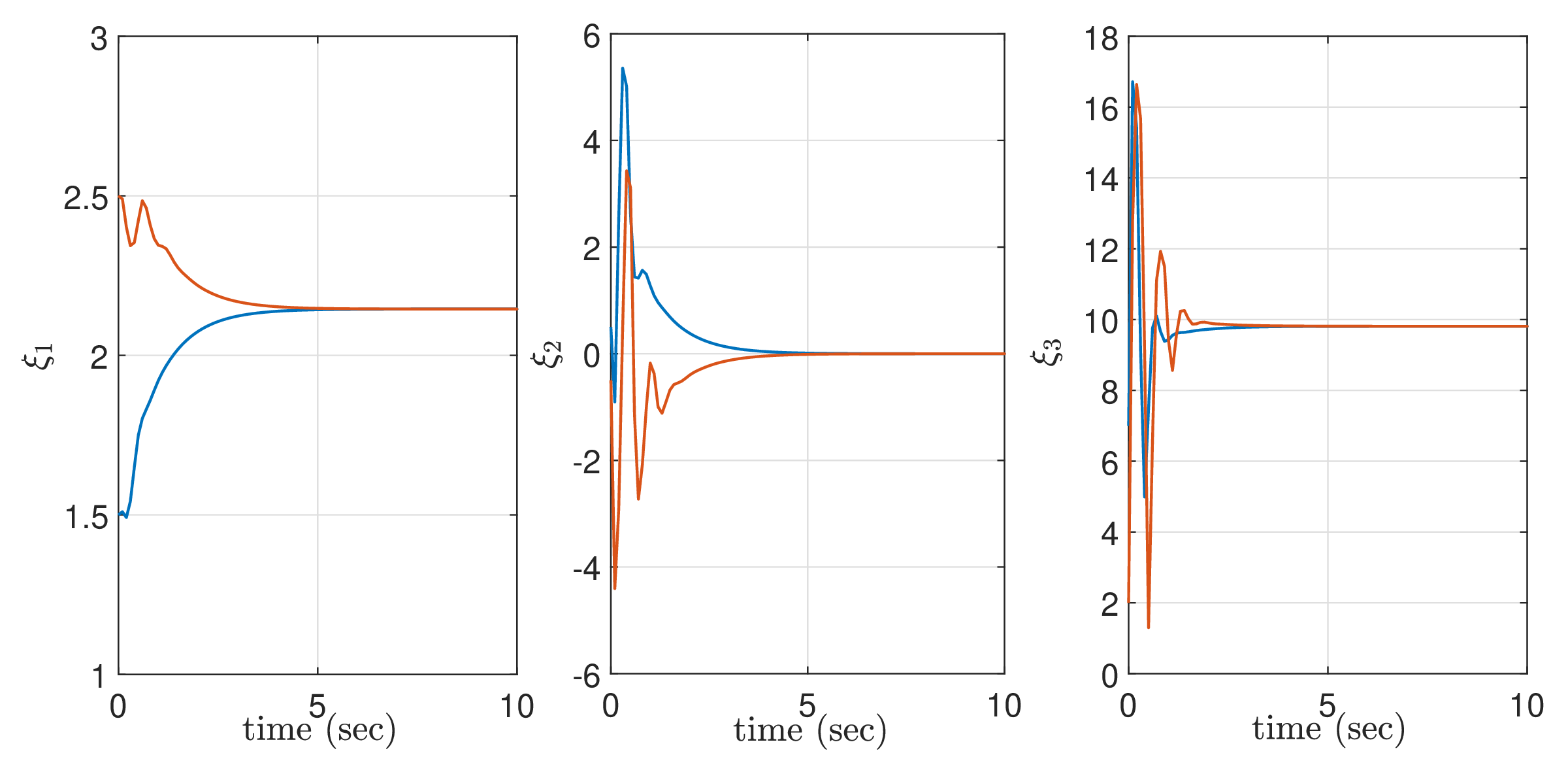}
\caption{Evolution of the states under a constant input $\hat{\upsilon}=200$ with the initial conditions $x_0=[1.5,0.5,7]$ (blue line) and $x_0=[2.5,-0.5,2]$ (red line).}
    \label{fig:levit_states}
\end{figure}

\begin{figure}\label{fig:levBound}
   \includegraphics[width=0.95\linewidth]{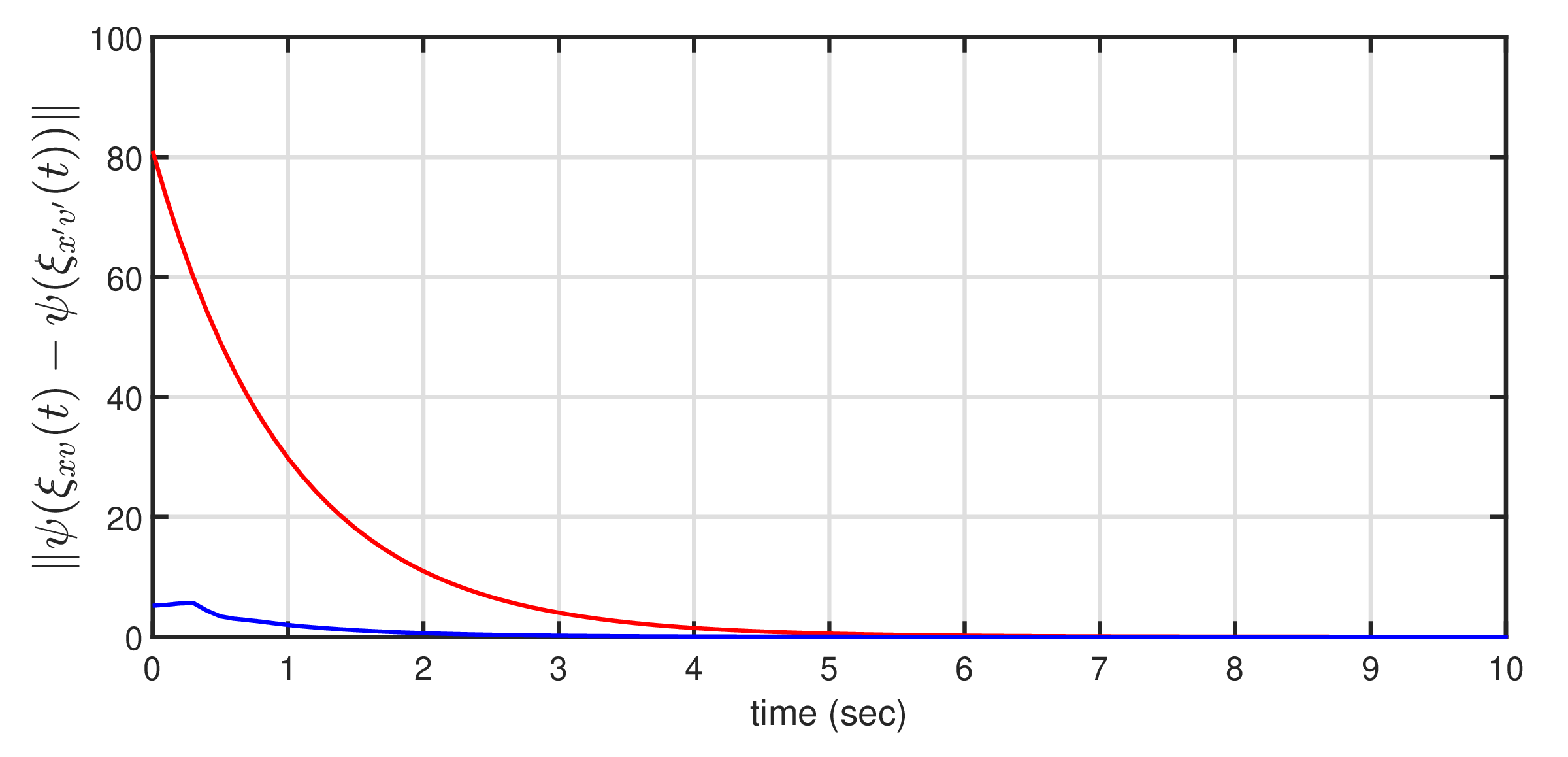}
    \caption{Distance between the trajectories (as computed using the distance metric in Theorem \ref{theorem:SFControl}, a.k.a closeness of trajectories) of the controlled system under a constant input $\hat{\upsilon}=200$ with the initial conditions $x=[1.5, 0.5, 7]$ and $x'=[2.5, -0.5, 2]$. The blue line denotes the closeness of the trajectories, and the red line is the probabilistic bound on the closeness, where c is computed with $\lVert\eta\rVert\lVert\Bar{\rho}\rVert=0.19$.
    }
    \label{fig:levit_bound}
\end{figure}

Using the learned mean and variance, we synthesized the controller based on the backstepping control scheme presented in Section \ref{Sec:control design}. We select the value of $\lambda_1 = 6.5$ and $\lambda_2 = 9.71$. Figures \ref{fig:levit_states} and \ref{fig:levit_bound} show the simulation results of the system (\ref{equ:magnetic_lev}) where $x_1$ is the angle of the manipulator link $\theta$ in radians and $x_2$ is the angular velocity $\dot\theta$ in rad/s. Figure \ref{fig:levit_states} shows the evolution of the system (\ref{equ:magnetic_lev}) under the control law (\ref{equ:controlaw}). The system starts from two different initial conditions, but it is obvious from the graphs that the states $x_1,x_2$ and $x_3$ converge towards each other in the case of both the initial states. Figure \ref{fig:levit_bound} shows the closeness of the trajectories of the system starting at different initial states. By closeness, we mean the distance between the trajectories computed based on the distance metric defined in Theorem \ref{theorem:SFControl}. The figure also shows the bounds on this closeness as defined in Theorem \ref{theorem:SFControl}. The `conservative bound' curve is the bound of the closeness with ${c}=5.318\times 10^{-5}$ computed with the value of $\lVert\Tilde{\eta}\rVert\lVert\Bar{\rho}\rVert=0.0049$. The probabilistic bound on the closeness is computed with $c=7.985\times10^{-6}$ computed based on the value of $\lVert\eta\rVert\lVert\Bar{\rho}\rVert=0.00188$. This clearly shows that the proposed control law renders the system (\ref{equ:magnetic_lev}) $\delta$-ISpS with respect to $\hat{\upsilon}$. Please note that the distance between the two trajectories converges to a non-zero value of $2.217\times 10^{-8}$.
\subsection{Two-link Manipulator}
We also conducted a similar experiment with a two-link manipulator \cite{2Regbook}
given by,
\begin{align}
        \dot \xi_1&=\xi_2,\nonumber \\
        \dot \xi_2 &= \underbrace{M^{-1}(\xi_1)\left[-H(\xi_1,\xi_2)-c(\xi_1)\right]}_{f(\xi)}+\underbrace{M^{-1}(\xi_1)}_{g(\xi)}\tau.\label{equ:2Rsys}
\end{align}Here,
\begin{gather*}
    M(x_1)=ml^2\begin{bmatrix}
    \left(\frac{5}{3}+\cos{\theta_2}\right) & \left(\frac{1}{3}+\frac{1}{2}\cos{\theta_2}\right)\\
    \left(\frac{1}{3}+\frac{1}{2}\cos{\theta_2}\right) & \frac{1}{3}
\end{bmatrix},\\
H(x_1,x_2) = ml^2\sin{\theta_2}\begin{bmatrix}
    -\frac{1}{2}\dot{\theta}_2^2-\dot{\theta}_1\dot{\theta}_2\\
    \frac{1}{2}\dot{\theta}_1^2
\end{bmatrix},\\
c(x_1)=ma_gl\begin{bmatrix}
    \frac{3}{2}\cos{\theta_1}+\frac{1}{2}\cos{(\theta_1+\theta_2)}\\
    \frac{1}{2}\cos{(\theta_1+\theta_2)}
\end{bmatrix},
\end{gather*}
$\xi=[\xi_1,\xi_2]^{\top}$, $\xi_1=\mathbf{p}$, $\xi_2=\dot{\mathbf{p}}$, $\mathbf{p}(t)=p=[\theta_1,\theta_2]^{\top}$, $\theta_1$ and $\theta_2$ are the angles of the two revolute joints and $a_g$ is the acceleration due to gravity.
We consider a compact set $\mathcal{X}=\left[-3,3\right]\times\left[-3,3\right]\times\left[-0.1,0.1\right]\times\left[-0.1,0.1\right]$. It is obvious that assumptions \ref{assume:RKHS}-\ref{assume:availabledata} hold for $f$. We train the unknown model $f$ using the Gaussian process with $400$ data samples of $x$ and $y=f(x)+w$, where $w\sim\mathcal{N}(0,\rho_f^2\mathbf{I}_2)$, $\rho_f=0.01$, collected by simulating the system with several randomly selected initial states. The considered kernel is $k_i(x,x')=\rho_{k_i}^2\exp{\left(\sum_{j=1}^4\frac{(x_j-x'_j)^2}{-2l^2_{ij}}\right)}$, $i=1,2$, where $\rho_{k_1}=178$, $\rho_{k_2}=287$, $l_{11}=2.11$, $l_{12}=0.516$, $l_{13}=190$, $l_{14}=356$, $l_{21}=2.28$, $l_{22}=0.494$, $l_{23}=142$ and $l_{24}=458$. Here, we abuse notation to represent $\theta_1,\theta_2,\dot{\theta}_1,\dot{\theta}_2$ as $x_1,x_2,x_3,x_4$ respectively. We computed the mean and variance as shown in (\ref{equ:mean}) and (\ref{equ:variance}) with $\lVert\Bar{\rho}\rVert=0.366$. For a value of $\lVert\eta\rVert\lVert\Bar{\rho}\rVert=0.19$, the probability interval is $[0.9803,0.9822]$ with a confidence of $1-10^{-10}$, $25^4$ realizations and ${c}=0.0767$.

\begin{figure}
 \includegraphics[width=\linewidth]{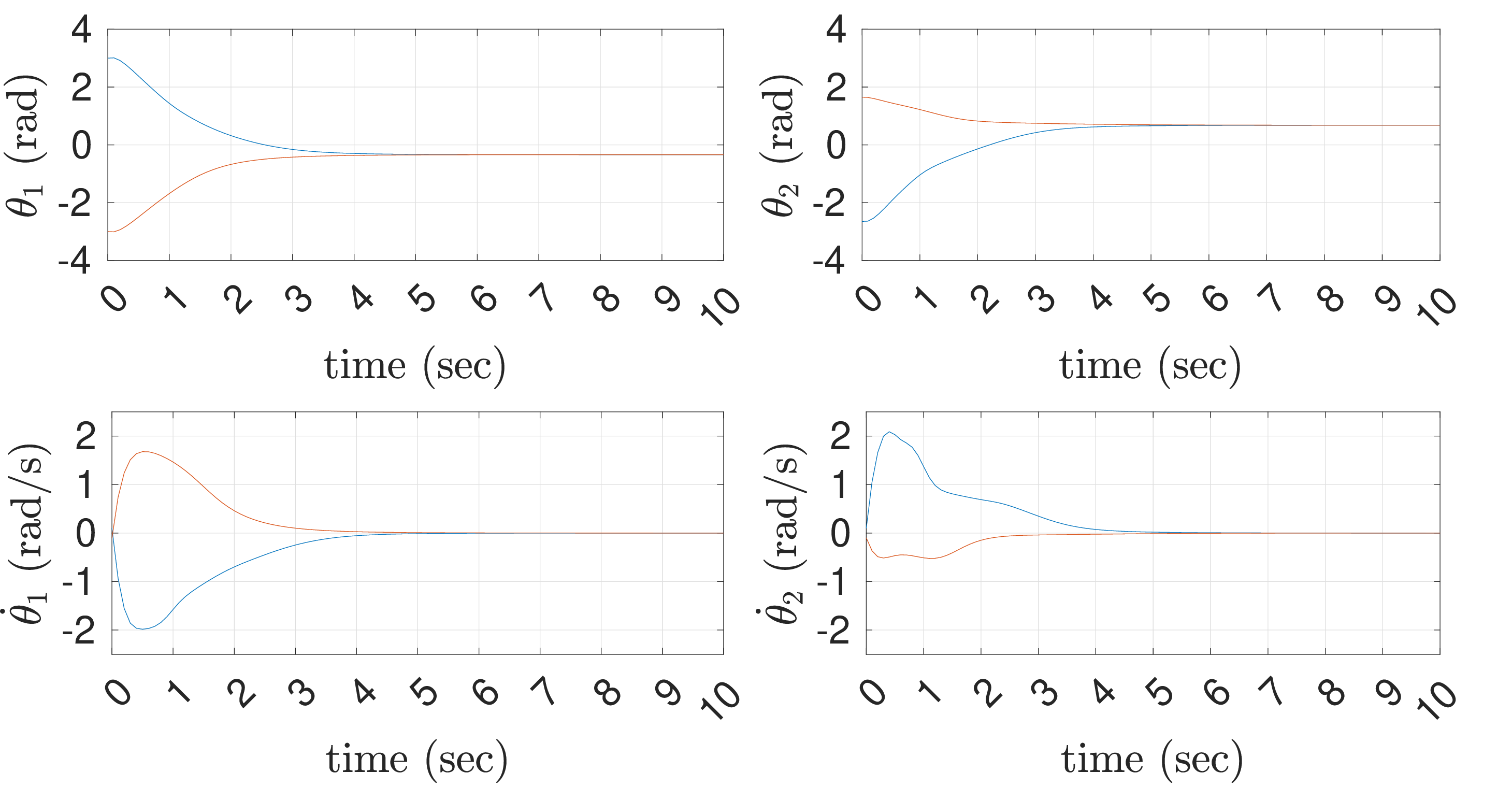}
\caption{Evolution of the system under a constant input $\hat{\upsilon}=[-1,2]$ with the initial conditions $x_0=[3,-2.65,1,1]$ (blue line) and $x_0=[-3,1.65,-1,-1]$ (orange line).}
    \label{fig:2Rtraj}
\end{figure}

\begin{figure}
    \includegraphics[width=\linewidth]{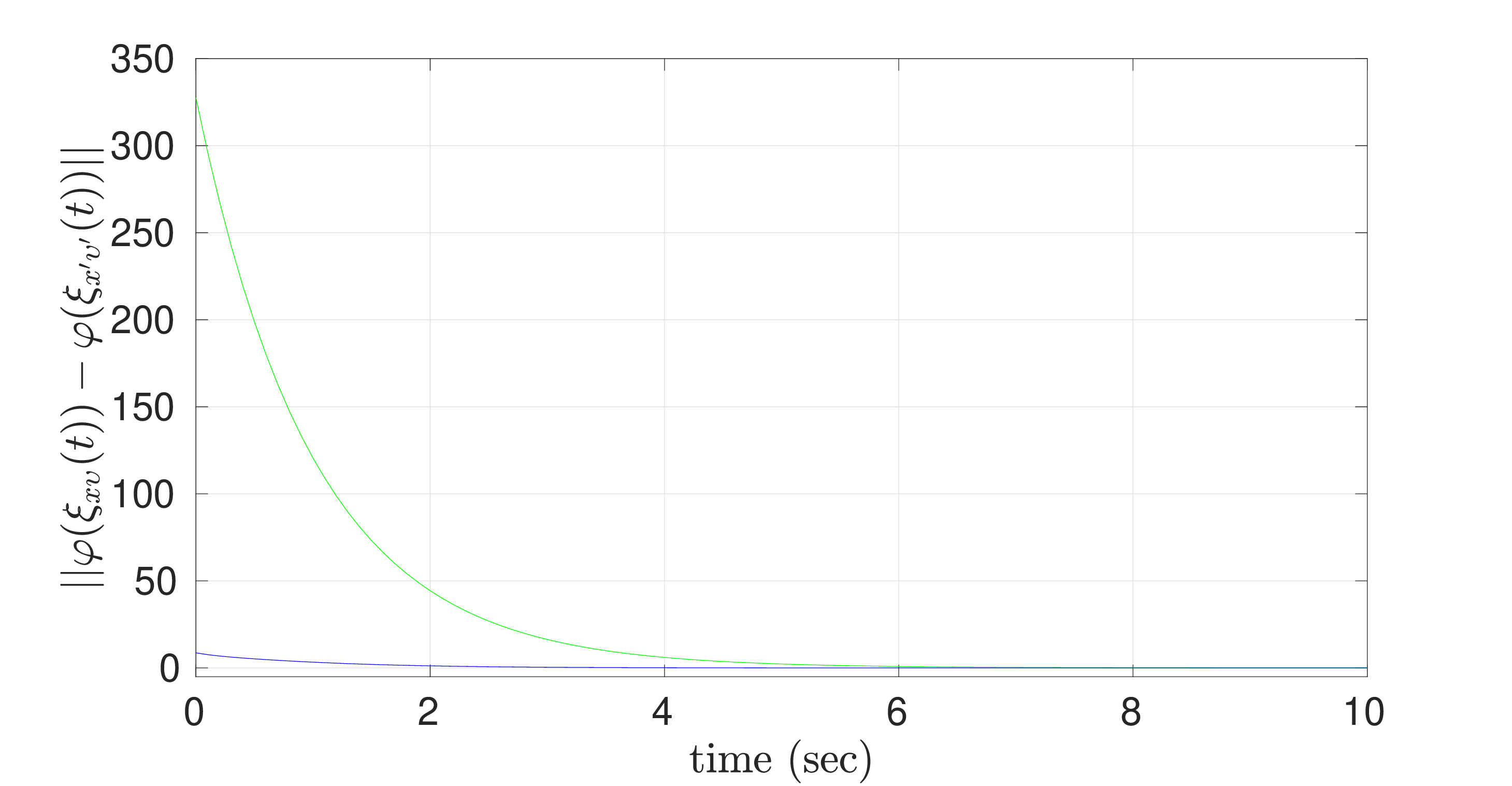}
    \caption{Distance between the trajectories (as computed using the distance metric in Theorem \ref{theorem:SFControl}, a.k.a closeness of trajectories) of the controlled system under a constant input $\hat{\upsilon}=[-1,2]$ with the initial conditions $x=[3,-2.65,0.1,0.1]$ and $x'=[-3,1.65,-0.1,-0.1]$. Here, the blue line denotes the closeness of the trajectories, and the green line is the probabilistic bound on the closeness, where c is computed with $\lVert\eta\rVert\lVert\Bar{\rho}\rVert=0.19$.}
    \label{fig:2Rdiff}
\end{figure}

We designed the controller as shown in Theorem \ref{theorem:SFControl} with the values of $\lambda_1 = 1$ and $\lambda_2=2.5$. Figure \ref{fig:2Rtraj} shows the evolution of the system (\ref{equ:2Rsys}) starting at two different initial conditions. It is obvious that the states converge towards each other. Figure \ref{fig:2Rdiff} shows the closeness of the trajectories and the bounds on the closeness for the system starting at two different initial conditions. This closeness converges to $2.44\times 10^{-8}$.
\section{Conclusion}
This paper introduces the concept and characterization of Incremental Input-to-Space practical Stability $\delta$-ISpS for unknown systems. It utilizes the Gaussian Process for learning the dynamics of a partially unknown system represented in strict-feedback form. The paper presented a control design approach using backstepping alongside $\delta$-ISpS Lyapunov functions aimed at synthesizing controllers to achieve $\delta$-ISpS for the system. Notably, this work marks the first attempt to synthesize a controller ensuring incremental stabilization, with a focus on probabilistic guarantees, for a class of partially unknown systems described in strict feedback form. To demonstrate the practicality of this approach, the paper implements the synthesized controllers using the proposed design scheme on two case studies, illustrating the convergence of system trajectories.
\bibliographystyle{ieeetr} 
\bibliography{sources}

\begin{thebibliography}{10}

\bibitem{synch}
A.~Hamadeh, G.-B. Stan, R.~Sepulchre, and J.~Goncalves, ``Global state synchronization in networks of cyclic feedback systems,'' {\em IEEE Transactions on Automatic Control}, vol.~57, no.~2, pp.~478--483, 2012.

\bibitem{synchComplex}
G.~Russo and M.~di~Bernardo, ``Contraction theory and master stability function: {L}inking two approaches to study synchronization of complex networks,'' {\em IEEE Transactions on Circuits and Systems II: Express Briefs}, vol.~56, no.~2, pp.~177--181, 2009.

\bibitem{synchOsci}
G.-B. Stan and R.~Sepulchre, ``Analysis of interconnected oscillators by dissipativity theory,'' {\em IEEE Transactions on Automatic Control}, vol.~52, no.~2, pp.~256--270, 2007.

\bibitem{modelCirc}
B.~N. Bond, Z.~Mahmood, Y.~Li, R.~Sredojevic, A.~Megretski, V.~Stojanovi, Y.~Avniel, and L.~Daniel, ``Compact modeling of nonlinear analog circuits using system identification via semidefinite programming and incremental stability certification,'' {\em IEEE Transactions on Computer-Aided Design of Integrated Circuits and Systems}, vol.~29, no.~8, pp.~1149--1162, 2010.

\bibitem{bisim1}
G.~Pola, A.~Girard, and P.~Tabuada, ``Approximately bisimilar symbolic models for nonlinear control systems,'' {\em Automatica}, vol.~44, no.~10, pp.~2508--2516, 2008.

\bibitem{bisim2}
R.~Majumdar and M.~Zamani, ``Approximately bisimilar symbolic models for digital control systems,'' in {\em Computer Aided Verification}, pp.~362--377, Springer Berlin Heidelberg, 2012.

\bibitem{zamani2017towards}
M.~Zamani, I.~Tkachev, and A.~Abate, ``Towards scalable synthesis of stochastic control systems,'' {\em Discrete Event Dynamic Systems}, vol.~27, pp.~341--369, 2017.

\bibitem{jagtap2020symbolic}
P.~Jagtap and M.~Zamani, ``Symbolic models for retarded jump--diffusion systems,'' {\em Automatica}, vol.~111, p.~108666, 2020.

\bibitem{jagtap2017quest}
P.~Jagtap and M.~Zamani, ``{QUEST: A} tool for state-space quantization-free synthesis of symbolic controllers,'' in {\em Quantitative Evaluation of Systems: 14th International Conference, QEST 2017, Berlin, Germany, September 5-7, 2017, Proceedings 14}, pp.~309--313, Springer, 2017.

\bibitem{characterize1}
W.~Lohmiller and J.-J.~E. Slotine, ``On contraction analysis for non-linear systems,'' {\em Automatica}, vol.~34, no.~6, pp.~683--696, 1998.

\bibitem{angeli}
D.~Angeli, ``A {L}yapunov approach to incremental stability properties,'' {\em IEEE Transactions on Automatic Control}, vol.~47, no.~3, pp.~410--421, 2002.

\bibitem{zamanicharacterize}
M.~Zamani, N.~{van de Wouw}, and R.~Majumdar, ``Backstepping controller synthesis and characterizations of incremental stability,'' {\em Systems \& Control Letters}, vol.~62, no.~10, pp.~949--962, 2013.

\bibitem{zamaninonsmooth}
M.~Zamani and N.~van~de Wouw, ``Controller synthesis for incremental stability: {A}pplication to symbolic controller synthesis,'' in {\em 2013 European Control Conference (ECC)}, pp.~2198--2203, 2013.

\bibitem{pushpakHamilton}
P.~Jagtap and M.~Zamani, ``Backstepping design for incremental stability of stochastic {H}amiltonian systems with jumps,'' {\em IEEE Transactions on Automatic Control}, vol.~63, no.~1, pp.~255--261, 2018.

\bibitem{deltaISS}
M.~Zamani and R.~Majumdar, ``A {L}yapunov approach in incremental stability,'' in {\em 2011 50th IEEE Conference on Decision and Control and European Control Conference}, pp.~302--307, 2011.

\bibitem{backsteppingzamani}
M.~Zamani and P.~Tabuada, ``Backstepping design for incremental stability,'' {\em IEEE Transactions on Automatic Control}, vol.~56, no.~9, pp.~2184--2189, 2011.

\bibitem{GPBook}
C.~K.~I. Williams and C.~E. Rasmussen, {\em Gaussian processes for machine learning}, vol.~2.
\newblock MIT press Cambridge, MA, 2006.

\bibitem{tracking}
T.~Beckers, D.~Kuli\'{c}, and S.~Hirche, ``Stable {G}aussian process based tracking control of {E}uler–{L}agrange systems,'' {\em Automatica}, vol.~103, p.~390–397, may 2019.

\bibitem{feedbackLinearization}
J.~Umlauft and S.~Hirche, ``Feedback linearization based on {G}aussian processes with event-triggered online learning,'' {\em IEEE Transactions on Automatic Control}, vol.~65, no.~10, pp.~4154--4169, 2020.

\bibitem{hircheControl}
J.~Umlauft, L.~Pöhler, and S.~Hirche, ``An uncertainty-based control {L}yapunov approach for control-affine systems modeled by {G}aussian process,'' {\em IEEE Control Systems Letters}, vol.~2, no.~3, pp.~483--488, 2018.

\bibitem{JagtapGP}
P.~Jagtap, G.~J. Pappas, and M.~Zamani, ``Control barrier functions for unknown nonlinear systems using {G}aussian processes,'' in {\em 2020 59th IEEE Conference on Decision and Control (CDC)}, pp.~3699--3704, 2020.

\bibitem{ISpS}
A.~Mironchenko, ``Criteria for input-to-state practical stability,'' {\em IEEE Transactions on Automatic Control}, vol.~64, no.~1, pp.~298--304, 2019.

\bibitem{GPBackStepping}
A.~Capone and S.~Hirche, ``Backstepping for partially unknown nonlinear systems using {G}aussian processes,'' {\em IEEE Control Systems Letters}, vol.~3, no.~2, pp.~416--421, 2019.

\bibitem{RKHS}
V.~I. Paulsen and M.~Raghupathi, {\em An Introduction to the Theory of Reproducing Kernel Hilbert Spaces}.
\newblock Cambridge Studies in Advanced Mathematics, Cambridge University Press, 2016.

\bibitem{strictFeedback}
M.~Krstic, P.~V. Kokotovic, and I.~Kanellakopoulos, {\em Nonlinear and Adaptive Control Design}.
\newblock USA: John Wiley \& Sons, Inc., 1st~ed., 1995.

\bibitem{existence}
E.~D. Sontag, {\em Mathematical Control Theory}.
\newblock Springer New York, NY, 1998.

\bibitem{deltaISSProof}
M.~Zamani, P.~M. Esfahani, R.~Majumdar, A.~Abate, and J.~Lygeros, ``Symbolic control of stochastic systems via approximately bisimilar finite abstractions,'' {\em IEEE Transactions on Automatic Control}, vol.~59, no.~12, pp.~3135--3150, 2014.

\bibitem{InfoGain}
N.~Srinivas, A.~Krause, S.~M. Kakade, and M.~W. Seeger, ``Information-theoretic regret bounds for {G}aussian process optimization in the bandit setting,'' {\em IEEE Transactions on Information Theory}, vol.~58, no.~5, pp.~3250--3265, 2012.

\bibitem{GPrandVar}
V.~V. Buldygin and Y.~V. Kozachenko, ``Sub-gaussian random variables,'' {\em Ukrainian Mathematical Journal}, vol.~32, pp.~483--489, Nov 1980.

\bibitem{bounds}
S.~R. Chowdhury and A.~Gopalan, ``On kernelized multi-armed bandits,'' in {\em Proceedings of the 34th International Conference on Machine Learning - Volume 70}, ICML'17, p.~844–853, JMLR.org, 2017.

\bibitem{adnaneDeterministic}
K.~Hashimoto, A.~Saoud, M.~Kishida, T.~Ushio, and D.~V. Dimarogonas, ``Learning-based symbolic abstractions for nonlinear control systems,'' {\em Automatica}, vol.~146, p.~110646, 2022.

\bibitem{jeltsema2009multidomain}
D.~Jeltsema and J.~M. Scherpen, ``Multidomain modeling of nonlinear networks and systems,'' {\em IEEE Control Systems Magazine}, vol.~29, no.~4, pp.~28--59, 2009.

\bibitem{ACMOpti}
C.~Zhu, R.~H. Byrd, P.~Lu, and J.~Nocedal, ``{Algorithm 778: L-BFGS-B: F}ortran subroutines for large-scale bound-constrained optimization,'' {\em ACM Trans. Math. Softw.}, vol.~23, p.~550–560, dec 1997.

\bibitem{2Regbook}
R.~M. Murray, S.~S. Sastry, and L.~Zexiang, {\em A Mathematical Introduction to Robotic Manipulation}.
\newblock USA: CRC Press, Inc., 1st~ed., 1994.

\end{thebibliography}

\end{document}